\documentclass[a4paper,twocolumn]{esapub} 
\usepackage{natbib}
\usepackage{epsfig}

\newcommand{\al}{\mbox{$^{26}$\hspace{-0.2em}Al}}

\newcommand{\pcmq}{\mbox{cm$^{-2}$}}
\newcommand{\psec}{\mbox{s$^{-1}$}}
\newcommand{\funit}{\mbox{ph \pcmq \psec}}
\def\deg{\hbox{$^\circ$}}

\title{SPI/INTEGRAL observation of 1809 keV gamma-ray line emission from 
       the Cygnus X region}
\author[1]{J. Kn\"odlseder}
\author[3,4]{M. Valsesia}
\author[1]{M. Allain}
\author[5]{S. Boggs}
\author[2]{R. Diehl}
\author[1]{P. Jean}
\author[2]{K. Kretschmer}
\author[1]{J.-P. Roques}
\author[2]{V. Sch\"onfelder}
\author[1]{G. Vedrenne}
\author[1]{P. von Ballmoos}
\author[1]{G. Weidenspointner}
\author[6]{C. Winkler}
\affil[1]{Centre d'\'Etude Spatiale des Rayonnements, B.P. 4346, 31028 
          Toulouse Cedex 4, France (knodlseder@cesr.fr)}
\affil[2]{Max-Planck-Institut f\"ur extraterrestrische Physik, Postfach 1312, 
          85741 Garching, Germany}
\affil[3]{IASF - CNR, via Bassini 15, 20133 Milan, Italy}
\affil[4]{Universit\`a di Pavia, Dipartimento di Fisica, via Bassi 6, 
	 	  27100 Pavia, Italy}
\affil[5]{SSL, University of California Berkeley, CA 94720, USA}
\affil[6]{ESA-ESTEC, Keplerlaan 1, 2201 AZ Noordwijk, The Netherlands}

\begin{document}


\maketitle

\begin{abstract}

We present first results on the observation of 1809~keV gamma-ray line 
emission from the Cygnus X region with the SPI imaging spectrometer. 
Our analysis is based on data from the performance verification phase 
of the INTEGRAL instruments and comprises 1.3 Ms of exposure time.
We observe a 1809 keV line flux of $(7.3 \pm 0.9) \times 10^{-5}$ \funit\ 
from a region delimited by galactic longitudes $73\deg$-$93\deg$ 
and $|b| \le 7\deg$ at a significance level of $8\sigma$.
The 1809 keV line appears moderately broadened, with an intrinsic FWHM of 
$3.3 \pm 1.3$ keV.
Although this broadening is only marginal (at the $2\sigma$ level our 
data are compatible with an unbroadened line), it could reflect 
the \al\ ejecta kinematics.

\end{abstract}

\section{Introduction}

The Cygnus X region is one of the most active nearby star forming regions 
in our Galaxy \citep{knodlseder03}.
It houses an exceptionally large number of O-type stars that enrich the 
interstellar medium with fresh nucleosynthesis products, either through 
their strong stellar winds, or eventually through their final supernova 
explosions. 
The radioactive isotope \al\ is a key tracer of this process, 
and the study of its 1809 keV decay radiation is a powerful probe for 
nucleosynthesis in this region.

COMPTEL measurements have revealed the presence of \al\ in Cygnus X 
\citep{delrio96,pluschke01}.
A deep study of the stellar populations in the Cygnus field suggests that 
most \al\ is produced in this region during hydrostatic nucleosynthesis in 
O-type and Wolf-Rayet stars; explosive nucleosynthesis in core collapse 
supernovae seems to play only a minor role.
Yet, the measured 1809 keV intensity is by at least a factor of 2 larger than 
expected from current nucleosynthesis models
\citep{knodlseder02}.
This may hint at an enhanced efficiency of hydrostatic \al\ nucleosynthesis,
such as expected for rotationally induced mixing in fast rotating 
O-type stars \citep{vuissoz04}.
Any nucleosynthesis process that is at work in Cygnus should in principle 
also hold for the entire Galaxy.
Hence Cygnus X may turn out to become the Rosetta stone for understanding 
\al\ nucleosynthesis galaxywide.
It is therefore particularly important to accurately measure the 
1809 keV gamma-ray line emission from this region, to precisely determine 
its intensity, to identify emission counterparts, and to study the line 
profile that holds information about \al\ ejecta dynamics.

\section{Analysed data}

The data that we analysed in this work were taken during the INTEGRAL 
performance verification phase.
They span the INTEGRAL orbits 14-25 (excluding orbit 24), covering 
1.3 Ms exposure time.
We analyse single-detector and double-detector data (SE and ME2).
Energy calibration has been performed orbit-wise, resulting in a relative 
(orbit-to-orbit) calibration precision of $\sim0.01$ keV and an absolute 
accuracy of $\sim0.05$ keV (Lonjou et al., these proceedings).

\begin{figure*}[!t]
  \center
  \epsfxsize=17cm \epsfclipon
  \epsfbox{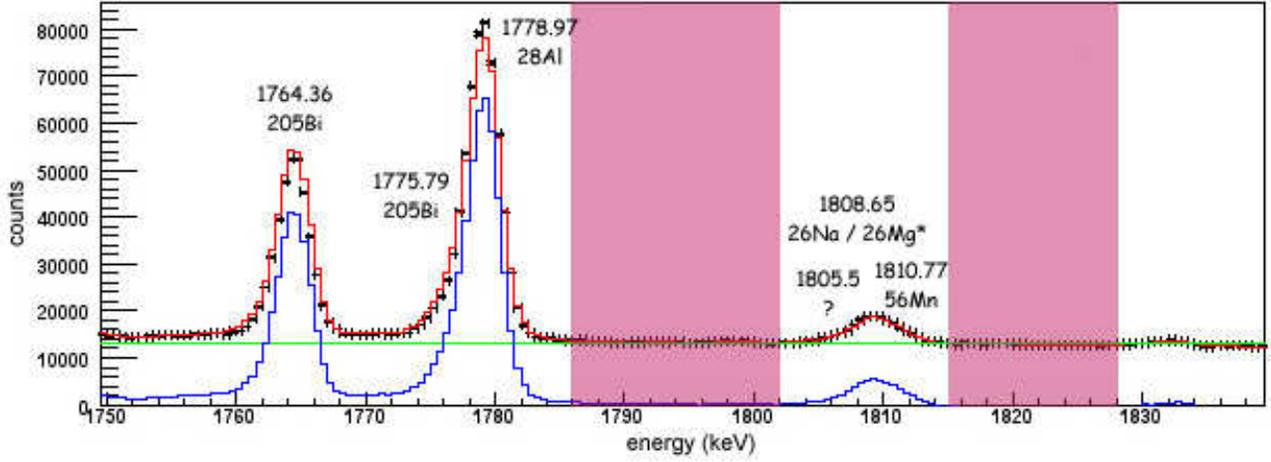}
  \caption{\label{fig:bgd}
  SPI single-detector spectrum in the area around 1809 keV
  together with the background model.
  (black/red histogram: line component; light/green line: continuum component; 
  grey/red histogram: summed model).
  The shaded areas indicate the energy bands that have been used for 
  modelling the continuum emission underlying the line. 
  Instrumental background lines have been labelled by their source isotopes. 
  }
\end{figure*}

\section{Background modelling}

The instrumental background of the SPI telescope at 1809 keV is 
composed of a broad line complex on top of a flat continuum 
distribution (cf.~Fig.~\ref{fig:bgd}).
The 1809 keV complex is a blend of at least 3 individual lines:
a yet unidentified component at about $1805.5 \pm 0.5$ keV,
a line at the \al\ rest energy of 1808.65 keV attributed to
$^{26}$Na($\beta^{-}$)$^{26}$Mg decay and to excitation of
$^{26}$Mg nuclei,
and a line at 1810.77 keV attributed to
$^{56}$Mn($\beta^{-}$)$^{56}$Fe decay
\citep{weidenspointner03}.
The relative contributions of each of the lines to the complex amount to 
roughly $10\%$, $60\%$ and $30\%$, respectively, and appear stable with 
time.

We model the energy and time dependence of the instrumental background 
using a two component model that allows for different time variabilities of 
line and continuum background.
The line component is predicted from empty field observations that are 
assumed to be free of celestial 1809 keV emission.
For this purpose, a total empty field exposure of 4.1 Ms has been collected 
by gathering all data for which SPI pointed at galactic latitudes 
$|b| > 20\deg$ (covering INTEGRAL orbits 14-106).
The spectra for this observation are summed for all detectors, and 
the line component is extracted by subtracting a constant offset whose 
level has been estimated from the adjacent energy bands 1786-1802 keV and 
1815-1828 keV.
The time dependent counting rate of the line component has been modelled 
by adjusting a multi-component background variation template to the 
counting rate history of the empty field data.
The components of the template reflect our current knowledge about the 
physical processes that give rise to the 1809 keV line complex
(see also Kn\"odlseder et al., these proceedings).

The continuum component was assumed to be constant in energy (note 
that we also used a constant to extract the line component from the 
empty field data, hence the line and continuum components are 
complementary and the sum of both should add without bias).
Its time dependent counting rate has been estimated by adjusting the 
germanium saturated event rate as activity tracer to the counting 
rate in adjacent energy bands (the same intervals have been used as for 
the extraction of the line component from the empty field data).
The adjustment has been performed using an adaptive running average that 
requires at least 10000 counts in the adjacent bands.
In this way, short term variabilities are explained by the activity 
tracer while long term variations come from the data themselves.

\begin{figure}[!t]
  \center
  \epsfxsize=8cm \epsfclipon
  \epsfbox{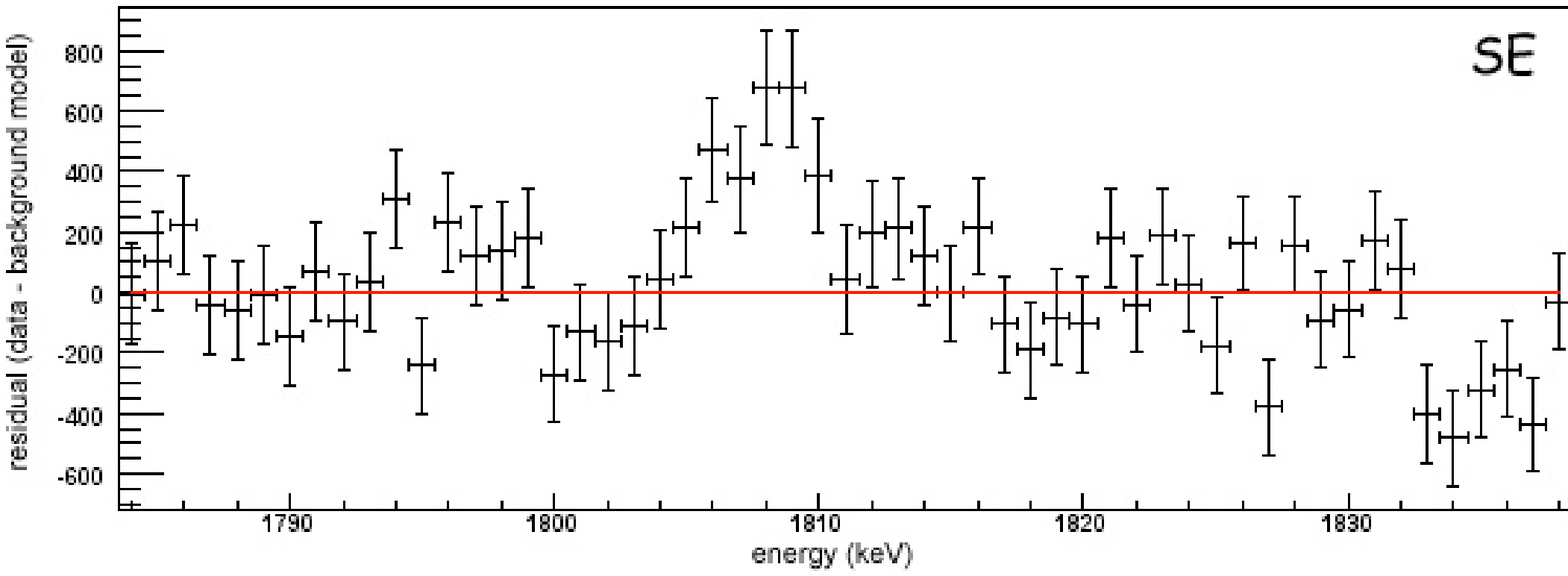}
  \center
  \epsfxsize=8cm \epsfclipon
  \epsfbox{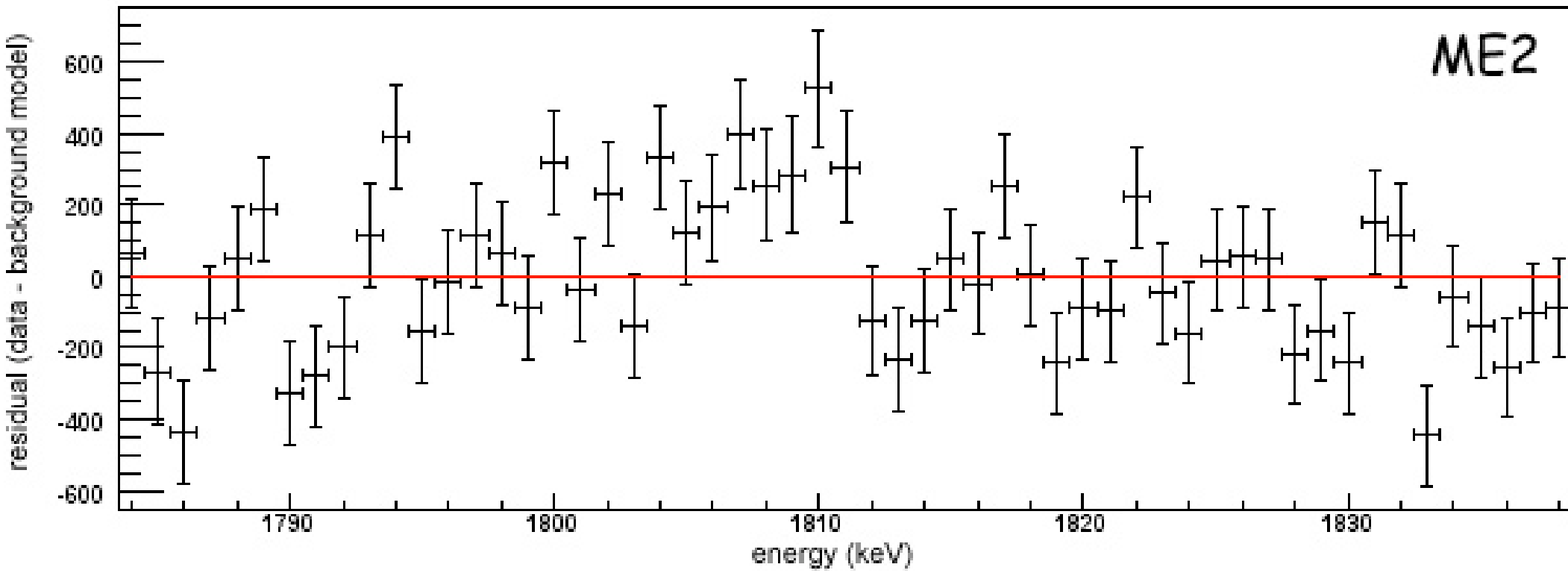}
  \caption{\label{fig:residuals}
  Residual SE and ME2 SPI energy spectra after subtraction of the background 
  model.
  For the SE spectrum a degradation correction has been 
  applied (section \ref{sec:adjustment}).
  }
\end{figure}

Figure \ref{fig:residuals} shows the background subtracted spectra 
for single-detector and double-detector events.
Clearly, a residual signal remains at 1809 keV in the spectra of both 
event types.

\section{Degradation correction}
\label{sec:adjustment}

\begin{figure*}[!t]
  \center
  \epsfxsize=17cm \epsfclipon
  \epsfbox{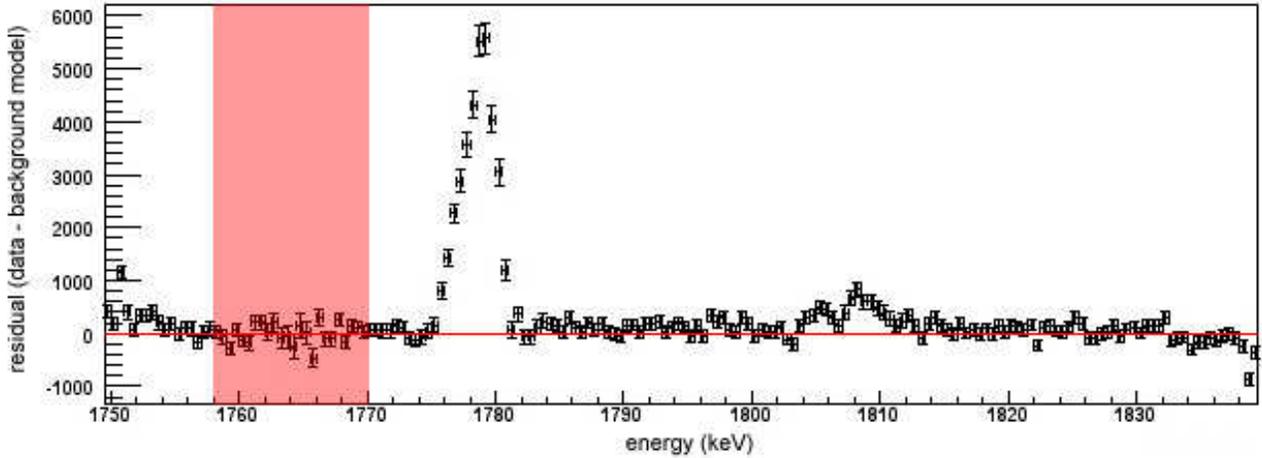}
  \caption{\label{fig:adjustment}
  SE residual spectrum after degradation correction, based on a $\chi^2$ 
  minimisation of the residuals in the energy 
  interval 1758-1770 keV (this energy band is indicated by a shaded 
  area).
  }
\end{figure*}

The exposure of the SPI camera to high-energy particle radiation in 
space leads to a degradation of the energy resolution with time due to
creation of trapping sites in the germanium crystals \citep{leleux03}.
These defects are regularly removed using an annealing procedure 
\citep{roques03}.
Since the empty field observations used for modelling the shape of the 
instrumental 1809 keV line complex were taken at difference epochs 
than the source data, and hence at different detector degradations, 
the shape of the instrumental line differs slightly between the 
source data and the background model.
A close look to Fig.~\ref{fig:bgd} reveals indeed that the background 
model shows broader lines than the source data
(detector degradation leads to an extension of the left wing of a 
line due to incomplete charge collection).

To match the spectral shape of source data and background model, we 
therefore reassign energies $E \to E'$ for each event using a random
number generator that follows the probability density function
\begin{equation}
 p(E'|E) = \frac{1}{\Delta E} e^{-(E-E')/\Delta E}
 \label{eq:explaw}
\end{equation}
where $E' \le E$ (i.e.~incomplete charge collection) and $\Delta E$ 
characterises the detector degradation.
Since Eq.~\ref{eq:explaw} effectively decreases the energy of each 
event (by $\Delta E$ on average) the background model has to be shifted 
to lower energies in order to match the data.
Also, a recalibration of the data is required after this correction, 
that we perform using the 1764.36 keV background line of $^{205}$Bi.

The degradation constant $\Delta E$ and the background shift were 
determined for each detector by minimising the residuals in the 
1764.36 keV background line.
This has been done using a $\chi^2$ minimisation applied to the energy 
interval 1758-1770 keV.
The resulting average degradation correction amounted to 
$\Delta E \approx 0.5$ keV, the average shift of the background model was
$-0.3$ keV.
The corrected SE residual spectrum is shown in Fig.~\ref{fig:adjustment}.
Note that the background model components were scaled by the 
fitting algorithm for $\chi^2$ minimisation and visual display since the 
normalisation of the 1764.36 keV line differs from those of the other 
background lines due to a different time history (without scaling the spectrum 
could never become flat).
For this reason a strong residual remains in Fig.~\ref{fig:adjustment} 
at 1779 keV and also the residual 1809 keV line signal is exaggerated.
For data analysis, however, no scaling has been applied to the background 
model.

Remaining residuals of the 1764.36 keV line are of the order of $1\%$ of 
the instrumental background line.
The expected celestial 1809 keV signal amounts to $\sim12\%$ of the 
counts in the 1809 keV complex. 
Assuming that the background subtraction leaves also $1\%$ residuals 
at 1809 keV, we estimate residuals from instrumental background that 
amount to $\sim10\%$ of the celestial 1809 keV signal.

\section{Spectral analysis}

\begin{figure*}[!t]
  \center
  \epsfxsize=8.4cm \epsfclipon
  \epsfbox{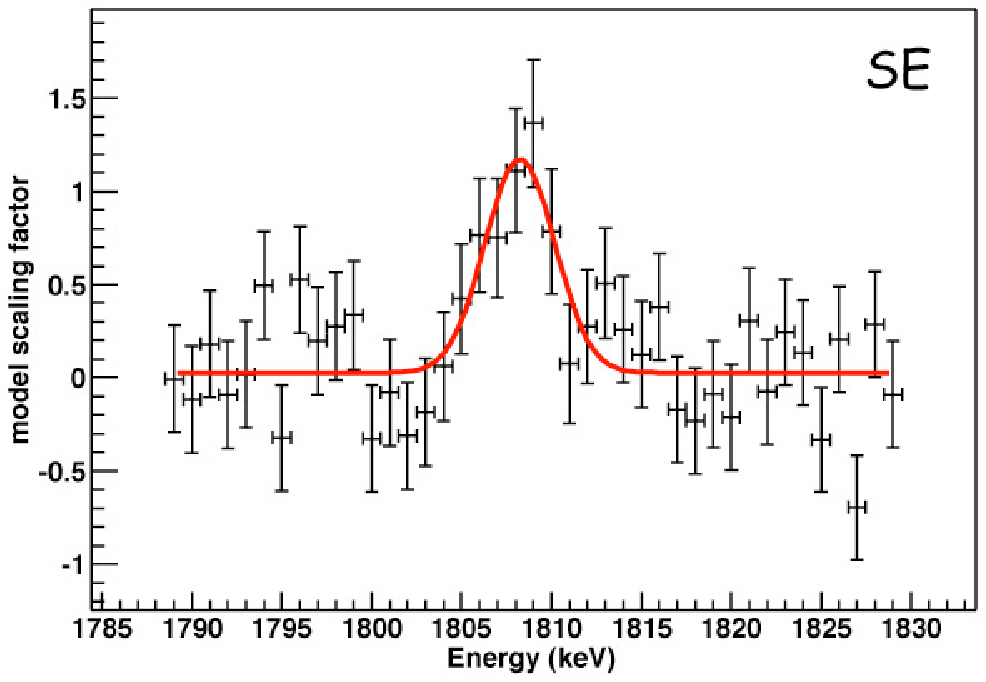}
  \hfill
  \epsfxsize=8.4cm \epsfclipon
  \epsfbox{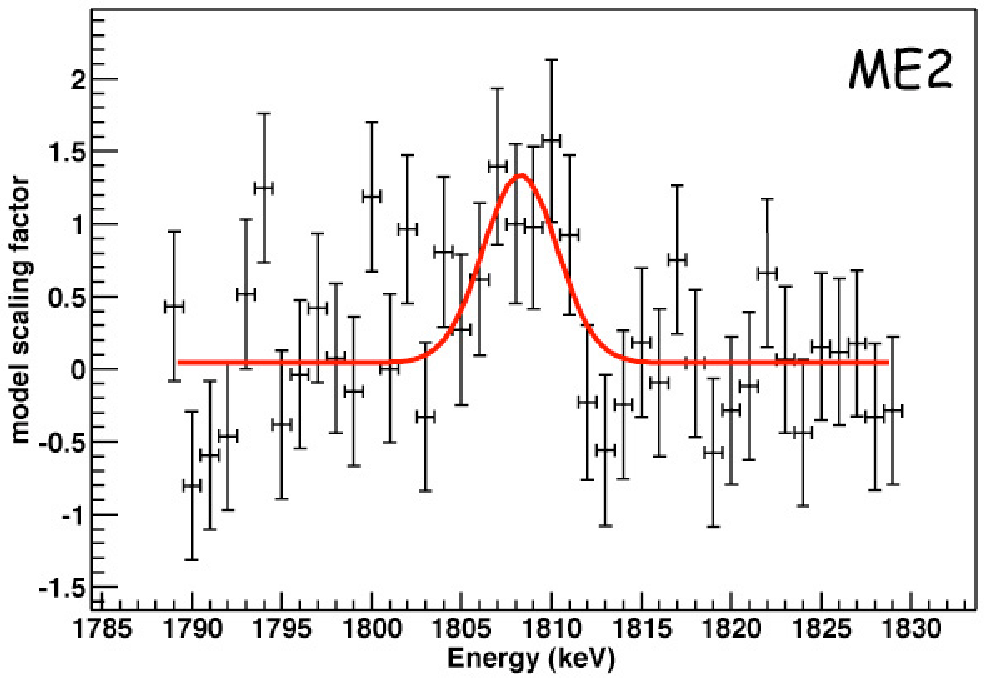}
  \caption{\label{fig:spectra}
  SPI SE and ME2 spectra of the Cygnus X region assuming an \al\ 
  line intensity distribution that follows the Dirbe 240 $\mu$m infrared 
  emission as a tracer of the massive star population.
  }
\end{figure*}

Straight background spectra subtraction does not make use of the coded-mask 
and field-of-view properties of SPI; on the other hand, image deconvolution 
is feasible only for strong signals. 
As we expect a weak signal, we derive spectra by fitting the 
amplitude of a given skymap of (expected) intensity distribution, per 1 keV 
wide energy bin. 
As skymap we used the Dirbe 240 $\mu$m intensity distribution, since this map 
has been shown to provide a reasonable tracer of galactic 1809 keV emission
\citep{knodlseder99}.

\begin{table}
\caption{Spectral analysis results.
         The flux is given in units of $10^{-5}$ \funit\ and has been 
         determined by integrating over galactic longitudes 
		 $70\deg$-$93\deg$ and laltitudes $b\le7\deg$. The energy and 
		 intrinsic line width are given in units of keV.}
\vspace{1em}
\begin{center}
\renewcommand{\arraystretch}{1.2}
\begin{tabular}{l c c c}
\hline 
Events & Flux & Energy & FWHM \\
\hline
SE  & $6.7 \pm 2.3$ & $1808.4 \pm 0.4$ & $3.2 \pm 1.7$ \\
ME2 & $8.1 \pm 3.1$ & $1808.4 \pm 0.7$ & $3.4 \pm 2.4$ \\
Sum & $7.2 \pm 1.8$ & $1808.4 \pm 0.3$ & $3.3 \pm 1.3$ \\
\hline
\end{tabular}
\label{tab:results} 
\end{center}
\end{table}

Figure \ref{fig:spectra} shows the resulting SE and ME2 spectra. 
Fitting Gaussian shaped lines on top of a constant provides 
flux, line position and line width estimates that are summarised in 
Table \ref{tab:results}.
The results for SE and ME2 have been combined by weighting with the 
statistical uncertainties.
The line position and width values are given after degradation correction 
and recalibration.
The line width is the intrinsic value after (quadratic) subtraction 
of the instrumental line widths.
The instrumental widths have been determined from the width of the 
1764.36 keV background line (after degradation correction and 
recalibration) which is known to be intrinsically narrow.
We obtain instrumental FWHMs of 3.2 and 3.5 keV for SE and ME2, 
respectively.

\section{Discussion and conclusions}

The measured 1809 keV line flux corresponds well to previous COMPTEL 
measurements.
From two years of COMPTEL data, \citet{delrio96} derived a flux of 
$(7.0 \pm 1.4) \times 10^{-5}$ \funit\ from a sky region covering
$l=[73\deg, 93\deg]$ and $b=[-7\deg,7\deg]$, corresponding to a 
detection significance of $5\sigma$.
Integrating the fitted Dirbe 240 $\mu$m map over the same area 
results in a SPI value of $(7.2 \pm 1.8) \times 10^{-5}$ \funit.
Our quoted flux uncertainty also includes the uncertainty in the 
line width.
Neglecting the line width uncertainty by analysing SPI data in single 
energy bands leads to a flux value of $(7.3 \pm 0.9) \times 10^{-5}$ \funit, 
corresponding to a SPI detection significance of $8\sigma$.

From the analysis of 9 years of COMPTEL data, \citet{pluschke01} 
derived a flux of $(10.3 \pm 2.0) \times 10^{-5}$ \funit\ from a 
slightly larger sky region covering $l=[70\deg, 96\deg]$ and 
$b=[-9\deg,25\deg]$.
For the same region we find a value of $(10.0 \pm 2.6) \times 10^{-5}$ 
\funit\ from the spectral analysis, and 
$(10.3 \pm 1.3) \times 10^{-5}$ from the single energy band analysis.
Thus, our flux measurements are consistent with earlier COMPTEL 
results, and although only 1.3 Ms of exposure time have been analysed 
so far, they present the most precise measurements of the 1809 keV line 
intensity from Cygnus~X that is available today.

SPI measured for the first time the precise energy of the 1809 keV 
line towards Cygnus, and the resulting value of $1808.4 \pm 0.3$ keV
is compatible with the rest energy of the \al\ line at 1808.65 keV.
When interpreted kinematically, the SPI measurement corresponds to a 
radial velocity of $-41 \pm 50$~km~s$^{-1}$ which is compatible with 
the radial velocity of $\sim~0$~km~s$^{-1}$ of massive star clusters 
found in Cygnus~X \citep{knodlseder03}.

\begin{figure}[!th]
  \center
  \epsfxsize=8cm \epsfclipon
  \epsfbox{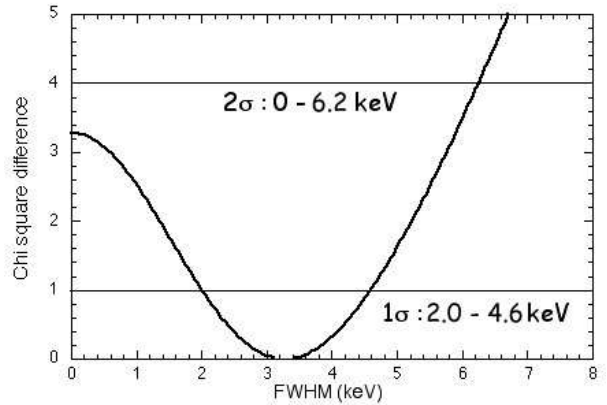}
  \caption{\label{fig:linewidth}
  $\chi^2$ of the combined SE and ME2 spectral fit as function of intrinsic 
  1809 keV line width.
  }
\end{figure}
 
SPI also provides for the first time an estimate of the intrinsic 
width of the 1809 keV line in Cygnus~X.
Figure \ref{fig:linewidth} shows the $\chi^2$ of the combined
fit of SE and ME2 spectra as function of the intrinsic line width.
No significant (i.e.~more than $3\sigma$) line broadening can be 
claimed from the present data, yet there are hints for a modest 
broadening.
The formal line width of $3.3 \pm 1.3$ keV corresponds to a Doppler 
broadening of $550 \pm 210$ km s$^{-1}$ (FWHM).
This translates into expansion velocities of 170-380 km s$^{-1}$ if
a thin expanding shell is assumed, or to 240-550 km s$^{-1}$ for
a homologously expanding bubble.

These values are relatively high with respect to expansion velocities of 
a few tens of km s$^{-1}$ that would have been expected from wind blown 
bubble expansion in Cygnus~X \citep{knodlseder03}.
Galactic rotation can also not explain this broadening since the 
velocity dispersion in this area of the Galaxy is quite small (also 
only a few tens of km s$^{-1}$).
Yet turbulent motions in hot superbubbles can reach velocities of a 
few 100 km s$^{-1}$ (De Avillez, private communication), and the \al\ 
ejecta may eventually follow these motions.
However, more observations of Cygnus X are required to confirm this 
hypothesis, by providing more stringent informations on the \al\ line 
profile.
More observations of Cygnus~X have been executed during the INTEGRAL 
AO-1 cycle, and further observations are scheduled for AO-2.
In the near future it can therefore be expected that SPI will provide
a more constraining measure of the \al\ line width, providing
valuable information on the \al\ ejecta dynamics and the ISM 
kinematics in the Cygnus~X region.


\end{document}